\documentstyle[twoside,fleqn,espcrc2,epsf]{article}

\newcommand{\AmS}{{\protect\the\textfont2
  A\kern-.1667em\lower.5ex\hbox{M}\kern-.125emS}}

\title{\vspace{-5.0cm}
\begin{flushright}
{\normalsize\tt RIKEN-AF-NP-234}\\
\vspace{-0.3cm}
{\normalsize\tt SNUTP-96-086}\\
\end{flushright}
\vspace*{2.5cm}
Quenched KS light hadron mass at \(\beta=6.5\)
on a \(64\times 48^3\) lattice\thanks{Poster
        presented at LATTICE96 by SO.  We thank the computation center
        of RIKEN for the use of VPP500/30.}}

\author{Seyong Kim\address{Center for Theoretical Physics,
        Seoul National University, Seoul, Korea}
        and Shigemi Ohta\address{Institute of Physical and Chemical
        Research (RIKEN), Wako-shi, Saitama 351-01, Japan}}
       
\begin{document}

\begin{abstract} 
We report our quenched staggered light hadron mass calculation at the
coupling of \(\beta = 6.5\) on a \(48^3 \times 64\) lattice, based on
an increased statistics of two hundred gauge configurations.
Staggered quark wall sources with mass of \(m_q a = 0.01, 0.005,
0.0025\) and \(0.00125\) are used.  Flavor symmetry is restored for
pion and \(\rho\) meson.  The lattice scale is estimated to be
\(a^{-1} = \mbox{3.7(2) GeV}\).
\end{abstract}

\maketitle

We report our light hadron mass calculation in quenched lattice
quantum chromodynamics (QCD) with an increased statistics of two
hundred gauge configurations.  Our lattice size is \(48^3 \times 64\)
and the inverse-squared coupling is \(\beta = 6.5\).  Staggered quark
wall sources of quark mass \(m_q a =\) 0.01, 0.005, 0.0025 and 0.00125
and point sink are used.  These parameters roughly correspond to a
physical box of \((\mbox{2.4 fm})^3\) and lattice cut off of \(a^{-1}
\sim \mbox{4 GeV}\) \cite{KO95}.

The space-like lattice size of 48 allows efficient use of a 24-node
partition of the RIKEN's 30-node VPP500/30 vector-parallel
supercomputer.  In generating the gauge configurations we use a
combination of a Metropolis update sweep followed by an
over-relaxation one.  The separation between two successive
hadron-mass calculations is 1000 such pairs of sweeps and take about 3
hours in total including the necessary disk accesses.  This separation
should be about equivalent with a series of earlier studies at lower
cutoff or smaller volume \cite{KS}.  With the current statistics of
200 configurations the autocorrelation in successive Nambu-Goldstone
pion propagators at time \(t = 20\) is about 15 \%.  All the
configurations used for the hadron-mass calculations, almost 2 Gbytes
each, are stored in a tape archive.  This will enable us to study
hadrons with strangeness and charm in the near future.  Further
details on our simulation method and characteristics were already
reported \cite{KO95,Ohta}.

We finished a covariant \(\chi^2\) analysis on our current 200 gauge
configurations (see Table \ref{tab:hadronmass}).  Results of two
fitting methods are summarized in the table.  Fit (1) uses the same
procedure as in the previous studies \cite{KS}: minimize the
correlated \(\chi^2\) function and choose the best fit by following
the maximum of (degrees of freedom) \(\times\) (confidence) \(/\)
(error).  Fit (2) uses the same quantity but choose from plateaus in
later time but before the signal disappears.  The reason why we
include these two fits are discussed in the following.

\begin{figure}
\vspace{9pt}
\epsfxsize=70mm
\leavevmode
\epsffile[50 50 410 302]{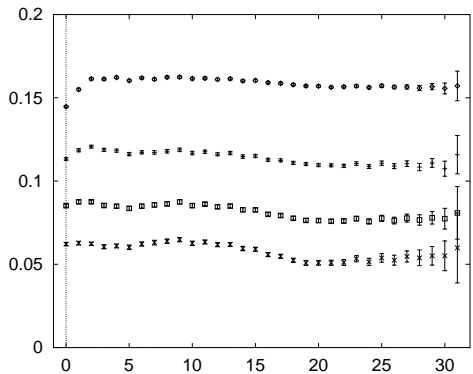}
\caption{Nambu-Goldstone pion effective mass at \protect\(\beta =
6.5\protect\) on \protect\(48^3 \times 64\protect\) lattice for quark
mass \protect\(m_q = 0.01\protect\), 0.005, 0.0025 and 0.00125.
Corner-wall source.  The errors are estimated by Jack-knife method.}
\label{fig:pieff}
\end{figure}

Let us look at the effective mass of Nambu-Goldstone pion plotted in
Figure \ref{fig:pieff}.
At a first glance, we observe nice long plateaus with small error
bars: if we neglect first four or five points in time, the remaining
points seem to align on a well defined plateau for each quark mass.
Indeed if we take weighted average of the effective mass from \(t=5\)
through 31, we get pion mass estimates of 0.1592(4), 0.1135(7),
0.0812(8) and 0.0583(9) for the four quark mass values of 0.01, 0.005,
0.0025 and 0.00125 respectively. However, closer inspection of Figure
\ref{fig:pieff} reveals strange wiggles and there seem to be two
plateaus for each quark mass: one for a higher mass in earlier time
and the other for a lower mass in later time.  This tendency is more
pronounced for lighter quark mass cases.  After we apply the same
procedure for globally fitting the pion propagator as in
ref. \cite{KS} ({\it ie\/} fit(1)), we are led to higher mass
estimates of 0.1618(4), 0.1170(5), 0.0856(8) and 0.0634(9) from the
best correlated \(\chi^2\) for \(t \le 14\).  If we choose lower
plateau region and neglect the first plateau ({\it ie\/} fit(2)), we
get estimates of 0.1575(3), 0.1109(4), 0.0767(7) and 0.052(1) for \(t
\ge 16\).

Both earlier and later plateaus may have problems.  The earlier one
can be contaminated by unwanted excited states.  The later one can be
dominated by noise, which usually lightens the fitted mass.  Or
perhaps the wiggles arise from intrinsic nature of effective mass
\cite{JLQCD95}.  Thus we are trying various different procedures in
extracting light hadron mass from the propagators.  In particular in
fit (3), we fit Jack-knife effective mass from farthest possible time
towards earlier time till the \(\chi^2\) begins to diverge.  As we
noted, fit (1) tends to give the earliest possible plateau in each
channel, and thus may suffer from unwanted excited state contribution.
Fit (2) tends to give later plateau and maybe free from such excited
state but noise may come in since we are far out in time.  Fit (3)
should be also free from excited states, but does not work well unless
the Jack-knife errors are well controlled.  With our current
statistics, fit (3) works at least for \(\pi\) and \(\pi_2\) and gives
their mass estimates that agree with fit (2).

\begin{figure}[t]
\vspace{9pt}
\epsfxsize=70mm
\leavevmode
\epsffile[50 50 410 302]{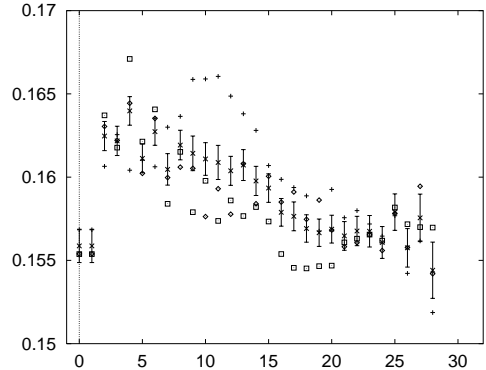}
\caption{Effective mass of pion calculated with 2000-sweep separation
for quark mass \protect\(m_q = 0.01\protect\).
\protect\(\Diamond\protect\) is the effective mass from the first 50,
\protect\(+\protect\) is from the second 50, \protect\(\Box\protect\)
is from the third 50, and \(\times\) is from all 150.}
\label{fig:pion.a1}
\end{figure}

We are also worried about whether the autocorrelation time is longer
than expected.  We are accumulating more statistics.  In addition to
the current 200 gauge configurations, we have so far accumulated 82
more configurations with twice larger separation (2000 pairs of
Metropolis and over-relaxed sweeps).  In a preliminary analysis, we
grouped our data into three 50-configuration sets. The current 200
configurations with 1000 sweep separations is reduced to 100
configurations by taking every other one and then divided into two,
earlier and later, 50-configuration groups.  We also take 50
configurations from 2000-sweep runs and make it the third group.  With
these three groups, we calculated effective mass for pions.  See
Figure \ref{fig:pion.a1} for the \(m_q = 0.01\) result.
We see the effective mass from each 50-configuration set fluctuates
around the value from the total 150 gauge configurations.  In
particular, the effective mass in the region typically selected by fit
(1) shows large fluctuations.  This suggests a longer autocorrelation
time and our earlier autocorrelation analysis of the pion propagator
could be misleading.  We are currently investigating this by
increasing statistics.
\begin{table*}
\setlength{\tabcolsep}{1.5pc}
\newlength{\digitwidth} \settowidth{\digitwidth}{\rm 0}
\catcode`?=\active \def?{\kern\digitwidth}
\caption{Hadron mass at \protect\(\beta = 6.5\protect\) on
\protect\(48^3 \times 64\protect\) lattice.}
\label{tab:hadronmass}
\begin{tabular*}{\textwidth}{lcrlrl}
\hline
\multicolumn{1}{l}{\protect\(m_qa\protect\)}&
particle &
\multicolumn{2}{c}{fit (1)} &
\multicolumn{2}{c}{fit (2)} \\
\hline\hline
0.01   &\protect\(\pi\protect\)   & 3-14&0.1618(4)&16-26&0.1575(3)\\%
       &\protect\(\pi_2\protect\) & 8-15&0.1594(6)&16-26&0.1604(6)\\%
       &\protect\(\sigma\protect\)& 6-11&0.314(3) & 8-15&0.318(3) \\%
       &\protect\(\rho\protect\)  &10-16&0.2451(9)&15-24&0.239(1) \\%
       &\protect\(\rho_2\protect\)&10-15&0.244(1) &14-23&0.239(1) \\%
       &\protect\(a_1\protect\)   &10-15&0.345(3) &14-23&0.343(6) \\%
       &\protect\(b_1\protect\)   &10-16&0.348(4) &15-24&0.37(1)  \\%
       &\protect\(N_1\protect\)   &10-17&0.364(2) &15-24&0.354(3) \\%
       &\protect\(N_2\protect\)   & 5-13&0.340(1) & 7-21&0.339(1) \\%
       &\protect\(\Delta\protect\)& 7-13&0.412(2) &11-17&0.404(3) \\%
\hline
0.005  &\protect\(\pi\protect\)   & 5-14&0.1170(5)&16-27&0.1109(4)\\%
       &\protect\(\pi_2\protect\) & 6-15&0.1151(7)&10-21&0.1152(8)\\%
       &\protect\(\sigma\protect\)& 6-15&0.311(6) &10-21&0.32(1)  \\%
       &\protect\(\rho\protect\)  &10-16&0.226(1) &15-24&0.218(2) \\%
       &\protect\(\rho_2\protect\)&10-16&0.222(2) &14-21&0.216(2) \\%
       &\protect\(a_1\protect\)   &10-16&0.320(5) &14-23&0.32(1)  \\%
       &\protect\(b_1\protect\)   &10-16&0.328(7) &12-27&0.33(1)  \\%
       &\protect\(N_1\protect\)   & 8-20&0.327(3) &10-20&0.321(3) \\%
       &\protect\(N_2\protect\)   & 5-13&0.301(2) & 7-14&0.298(2) \\%
       &\protect\(\Delta\protect\)& 6-13&0.394(2) &12-17&0.372(5) \\%
\hline
0.0025 &\protect\(\pi\protect\)   & 8-14&0.0856(8)&18-23&0.0767(7)\\%
       &\protect\(\pi_2\protect\) & 5-20&0.086(1) &14-21&0.0823(2)\\%
       &\protect\(\sigma\protect\)& 4-20&0.294(7) & 7-14&0.32(2)  \\%
       &\protect\(\rho\protect\)  & 7-16&0.222(2) &10-16&0.218(2) \\%
       &\protect\(\rho_2\protect\)& 8-19&0.217(2) &10-21&0.211(3) \\%
       &\protect\(a_1\protect\)   & 8-14&0.315(5) &10-19&0.306(8) \\%
       &\protect\(b_1\protect\)   & 7-16&0.326(7) &10-25&0.32(1)  \\%
       &\protect\(N_1\protect\)   & 7-20&0.317(4) & 8-20&0.308(5) \\%
       &\protect\(N_2\protect\)   & 4-15&0.281(2) & 7-15&0.277(3) \\%
       &\protect\(\Delta\protect\)& 5-10&0.394(3) & 8-14&0.370(5) \\%
\hline
0.00125&\protect\(\pi\protect\)   & 8-13&0.0634(9)&18-23&0.052(1) \\%
       &\protect\(\pi_2\protect\) & 3-13&0.066(2) & 5-13&0.065(3) \\%
       &\protect\(\sigma\protect\)& 4-13&0.28(1)  & 7-14&0.32(3)  \\%
       &\protect\(\rho\protect\)  & 6-16&0.219(2) &10-25&0.215(4) \\%
       &\protect\(\rho_2\protect\)& 4-10&0.227(3) & 7-25&0.217(4) \\%
       &\protect\(a_1\protect\)   & 4-10&0.317(4) & 8-20&0.302(9) \\%
       &\protect\(b_1\protect\)   & 5-17&0.321(7) & 6-16&0.306(8) \\%
       &\protect\(N_1\protect\)   & 4-20&0.318(5) & 6-23&0.314(7) \\%
       &\protect\(N_2\protect\)   & 4-15&0.271(4) & 5-13&0.272(4) \\%
       &\protect\(\Delta\protect\)& 4-22&0.403(5) & 6-25&0.376(5) \\%
\hline
\end{tabular*}
\end{table*}

\begin{figure}[t]
\vspace{9pt}
\epsfxsize=70mm
\leavevmode
\epsffile[50 50 410 302]{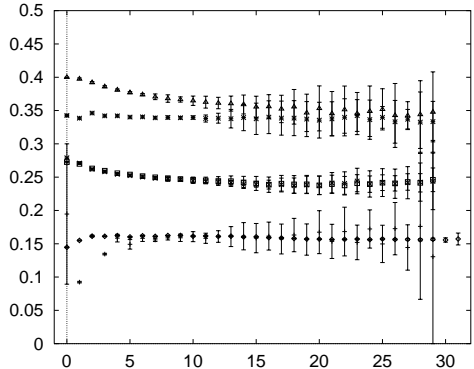}
\caption{Effective mass of flavor symmetry partners at \protect\(\beta
= 6.5\protect\) on \protect\(48^3 \times 64\protect\) lattice for
quark mass \protect\(m_q = 0.01\protect\): \protect\(N_1\protect\) and
\protect\(N_2\protect\) (top), \protect\(\rho\protect\) and
\protect\(\rho_2\protect\) (middle) and \protect\(\pi\protect\) and
\protect\(\pi_2\protect\) (bottom).  All show flavor symmetry
restoration.}
\label{fig:0.01}
\end{figure}

In Figure \ref{fig:0.01} we plot effective mass of flavor symmetry
partners: Nambu-Goldstone pion \(\pi\) and non-Goldstone one
\(\pi_2\), \(\rho\) and \(\rho_2\), and \(N_1\) and \(N_2\), for the
heaviest quark mass value of 0.01.
We clearly observe that \(\pi\) and \(\pi_2\) are on top of each other,
and so are \(\rho\) and \(\rho_2\). The same is observed for pions
and \(\rho\) mesons for the lighter quark mass values, 0.005, 0.0025
and 0.00125, albeit with more noise. From this we conclude that the
flavor symmetry is restored in the present calculation.
We also observe that \(N_2\) signal, from the ``even-point-wall''
source, is nearly flat and \(N_1\), from the ``corner-wall'' source,
seems to converge with it for large \(t\).  Thus we will use \(N_2\)
for nucleon mass estimation.

\begin{figure}[t]
\vspace{9pt}
\epsfxsize=70mm
\leavevmode
\epsffile[98 222 452 549]{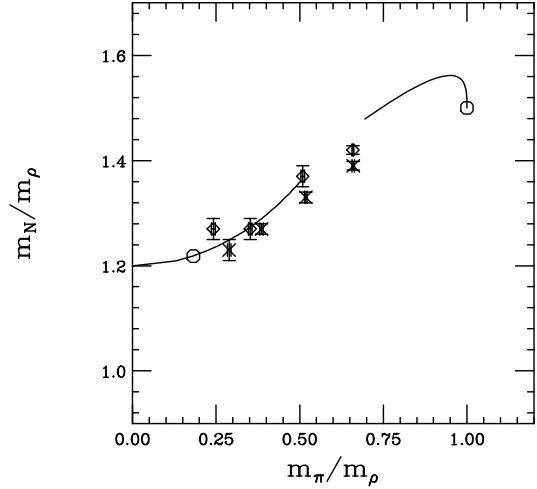}
\caption{Edinburgh plot at \protect\(\beta = 6.5\protect\) on
\protect\(48^3 \times 64\protect\) lattice for quark mass
\protect\(m_q = 0.01\protect\), 0.005, 0.0025 and 0.00125.  The two
different fits are plotted for each quark mass value.}
\label{fig:edinburgh}
\end{figure}

Figure \ref{fig:edinburgh} gives the Edinburgh plot in which we
included the data from fits (1) and (2).
The errors in the figure are obtained by assuming that the relative
error in each quantity is independent of each other.

Conclusions.  We are close enough to the continuum to see flavor
symmetry restored for both \(\pi\) and \(\rho\).  The lattice scale
estimated by \(\rho\) mass at zero quark mass, \(m_\rho(m_q=0) =
0.21(1)\), is \(a^{-1} = \) 3.7(2) GeV.  An interesting Edinburgh plot
is obtained, with \(m_\pi/m_\rho\) as small as 0.25 and \(m_N/m_\rho\)
within the error bar of the experimental value.  We still do not
understand the systematic error associated with plateau selection in
hadron effective mass.  We plan to accumulate more statistics to study
this.  The answer to the question on the quenched chiral log in the
pion mass \(m_\pi\) and chiral condensate \(\langle{\bar
\psi}\psi\rangle\) \cite{Sharpe,Golt_Bernard} should wait until we
sort out this systematic error.


\end{document}